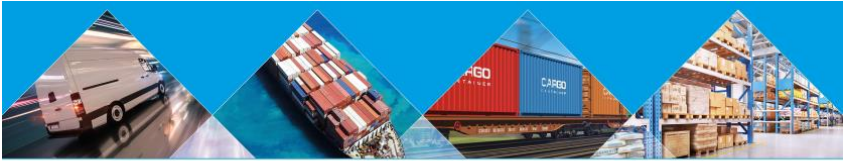

# Modeling and analysis of alternative distribution and Physical Internet schemes in urban area

Hao Jiang[1], Eric Ballot[2] and Shenle Pan[3]

MINES ParisTech, Paris, France

Corresponding author: iris.jiang_hao@mines-paristech.fr

*Abstract: Urban logistics is becoming more complicated and costlier due to new challenges in recent years. Since the main problem lies on congestion, the clean vehicle is not necessarily the most effective solution. There is thus a need to redesign the logistics networks in the city. This paper proposes a methodology to evaluate different distribution schemes in the city among which we find the most efficient and sustainable one. External impacts are added to the analysis of schemes, including accident, air pollution, climate change, noise, and congestion. An optimization model based on an analytical model is developed to optimize transportation means and distribution schemes. Results based on Bordeaux city show that PI scheme improves the performances of distribution.*

## 1 Introduction

Urban logistics, indispensable to the economy, has always been an issue of crucial importance due to continuously increasing freight volume. It is arduous to keep urban logistics reliable, affordable and environmentally friendly.

Some recent trends indicate that urban logistics is becoming even more complicated and costly. For example, the urban population has been growing by leaps and bounds in recent decades. It contributes to 55% of the world's population in 2018, and the proportion is expected to increase to 68% by 2050 reported by the United Nations (2018). Undoubtfully the urbanization and the phenomenon of mega-cities have brought more urban logistics challenges. Moreover, e-commerce is under drastic development. 1.042 billion orders were generated of T-mall, a Chinese e-commerce platform, during a single day on 11th November 2018 ("Alibaba Singles Day 2018: Record sales on largest shopping event day," n.d.). Transporting an enormous amount of goods like this directly to customers instead of retail stores results in a rising number of freight movements in the city. Furthermore, the higher service level offered in terms of lead time puts pressure on carriers. Some of them start to provide same-day delivery or even 2-hours delivery (e.g., Amazon Prime Now) in recent years. From a logistical point of view, high-speed delivery will further increase freight movements, which will add more challenges to direct-to-customer distribution in the city. Additionally, the rise of sharing economy could provide opportunities for improving the performances of urban logistics, but also makes it more complex to consolidate at the same time. Last but not least, more and more attention has been paid to sustainability and externalities. Accidents, air pollution, climate change, noise and congestion caused by urban freight transport are affecting the life of citizens and the global climate in a negative way.

In order to alleviate the negative impacts from economic, environmental and social aspects, a multitude of solutions has been proposed at an operational, tactical and organizational level.

These include new transportation means (electric freight vehicles, cargo cycles or drones), public policy and infrastructure (limited traffic zones, loading/unloading area, multi-use lanes, pick-up points, and off-hour deliveries), multi-echelon networks (*Urban Consolidation Center (UCC), Physical Internet (PI)* hubs), and also optimization of transport management, routing, etc. (Cleophas et al., 2018; Macharis and Kin, 2017; Ranieri et al., 2018). Since the distribution operations such as *Vehicle Routing Problem (VRP) and* depot location problem have already been well studied, our research will be focused on the organizational level and the distribution network design in urban area.

Based on current studies, the objective of this paper is to analytically explore the economic, environmental and social performances of different logistics schemes in the city to improve the efficiency and sustainability. Thus, an analytical model based on the continuous approximation methodology is developed to evaluate the performances of different logistics schemes, considering the transportation cost and external impacts (accidents, air pollution, climate change, noise, and congestion). The impact of major factors (e.g., vehicle speed, lead time, shipment size) of the design of logistics networks is further studied. In addition, an optimization model is developed to study the impact of transportation means on distribution performance. The models are validated through a case study of six suppliers from different sectors in Bordeaux, France. The case study was made possible by Club Déméter.

The structure of the paper is as follows. A literature review is conducted on solutions to urban logistics problems, cost models of urban logistics and external impact parameters in the next section. The mathematical models are illustrated in Section 3. Then section 4 describes the logistics schemes of the case study in which data is from six sectors including fast consumer moving goods, food, restaurants and cosmetics. Section 5 shows the results of the analysis. After a discussion on the results and the models (Section 6), this paper goes to the end with conclusions and future research (Section 7).

## 2 Literature review

### 2.1 Solutions to urban logistics problems

Freight transport is to transport the right goods at the right time to the right place at a low cost. Nowadays, it becomes even more challenging to manage due to the demand of customers for short lead-time delivery and their increasing attention to sustainability. Meanwhile, challenges of freight transport are undoubtfully hard to be dealt with in the context of urban area with the highest traffic density. In the literature, numerous studies have been seeking possible solutions to improve the efficiency and sustainability of distribution in the city.

#### 2.1.1 Multi-echelon networks

Among new urban logistics strategies, Urban Consolidation Center (UCC) has received increasing academic and practical attention. UCC is most commonly defined as a logistics facility that is situated in relatively close proximity to the geographic area that it serves a city center, an entire town or a specific site (e.g. shopping center), from which consolidated deliveries are carried out within that area (Michael Browne et al., 2005). It is found that the monetary and environmental benefits of UCC are either owing to better utilization of vehicle capacity by consolidating the goods before entering the city, or because of cheaper storage space at the UCC. Furthermore, the possibility of improving efficiency and sustainability highly depends on the characteristics of the city, location of the UCC and other factors. (Browne et al., 2011; Lin et al., n.d.; Michael Browne et al., 2005).

Another interesting concept related to multi-echelon networks is the Physical Internet (PI), an open global logistics system founded on physical, digital and operational inter-connectivity through encapsulation, interfaces and protocols (Montreuil et al., 2012). PI aims at reverting the huge unsustainability of the current way we transport, handle, store, realize, supply and use physical objects around the world (Ballot et al., 2014). The idea is to put all the resources (vehicles, PI hubs, PI containers…) in an interconnected distribution network into best use by sharing resources.

### 2.1.2  Dynamic delivery system

Short lead time required by customers such as same-day delivery leads to new optimization challenges since the same-day delivery makes the operations more dynamic. Although a multitude of researchers has contributed to dynamic vehicle routing problems (Berbeglia et al., 2010; Pillac et al., 2013; Psaraftis et al., 2016), it is still challenging to deal with the deliveries of more uncertain orders. Within shorter lead time, several short trips of delivering fewer orders work better than a long trip delivering all orders. Initiatives like crowd shipping then arose. A study of *Vehicle Routing Problem with Occasional Drivers (VRPOD)* proves that the compensation scheme for occasional drivers has a significant impact on the cost-effectiveness of crowd shipping (Archetti et al., 2015).

Our research will be focused on urban logistics schemes regarding the multi-echelon networks and transportation means applied for delivery. Not only because the former is one of the most basic study fields since the last century, but also it improves the urban logistics problem at an organizational level. Furthermore, it is interesting to investigate the optimization of transportation means to validate the trend of lighter duty vehicles usage.

## 2.2  Logistics cost models

Various logistics cost models have been developed in the literature. The very first one should be the approach to estimate the length of the vehicle routing problem (VRP) raised by Daganzo (2005). A cost model of FMCG company is developed, introducing a way to determine the number of tours. As a starting point, only the volume restriction is used to determine the number of tours. Few iterations are needed when assuming that the shift length of the driver is approximately matched with the volume capacity restriction (Kin et al., 2018). The cost of a product is defined as a function of inventory, handling and transportation costs. Detailly speaking, inventory cost is a function of replenishment frequency, order size, standard deviation of demand, standard deviation of the supply, transport time, interest rate and value of the goods transported; the packaging density is regarded as an independent variable of handling cost; the transport cost is a function of the distance, shipment size and frequency, value density, mode, speed and the reliability of the mode used. (Crainic et al., 2004; Cuda, 2015; Hemmelmayr and Gabriel, 2012) The application of transshipment points results in multi-echelon networks and related costs (Cuda, 2015).

## 2.3  External impacts parameters

The external impacts such as accidents, air pollution, climate change, noise, and congestion should be included when redesigning the logistics networks in the city since growing attention has been poured into the sustainability of urban logistics. Yet there are very few surveys reporting the external impacts parameters because it requires a large amount of human and material resources. Among those limited resources, it's scarcely possible to have a standard admitted by the majority due to the nonnegligible differences from each other.

*Table 1: External impacts parameters from four reports*

|  | TU Delft<br>(€/1000t.km) | INFRAS/IWW<br>(€ct/v.km) |
|---|---|---|
| Accident cost | 10.2 | 3.4 |
| Air pollution cost | 6.7 | 20.5 |
| Climate change cost | High scenario 9.8<br>Low scenario 1.7 | 6.3 |
| Noise cost | 1.8 | 27.4 |
| Congestion cost | 4 (€/v.km) | NA |

Among the two units of t.km and v.km, we decide to apply the v.km. Because t.km is normally used for macroscopic analysis, not for small-scale delivery analysis. Although empty runs deliver nothing, they generate external impacts as well. For calculation according to v.km, we apply the data from INFRAS/IWW and congestion cost from TU Delft.

According to the literature review, urban logistics schemes regarding the multi-echelon networks and transportation means applied for delivery are designed. Furthermore, the calculation of external impacts is added to the cost model of different logistics schemes, which could be a contribution to the literature of the logistics cost model.

## 3 Methodology

### 3.1 General methodology

To solve the research problem of inefficiency and unsustainability in urban logistics, the research emphasis of this paper lies on distribution schemes design at the organizational level.

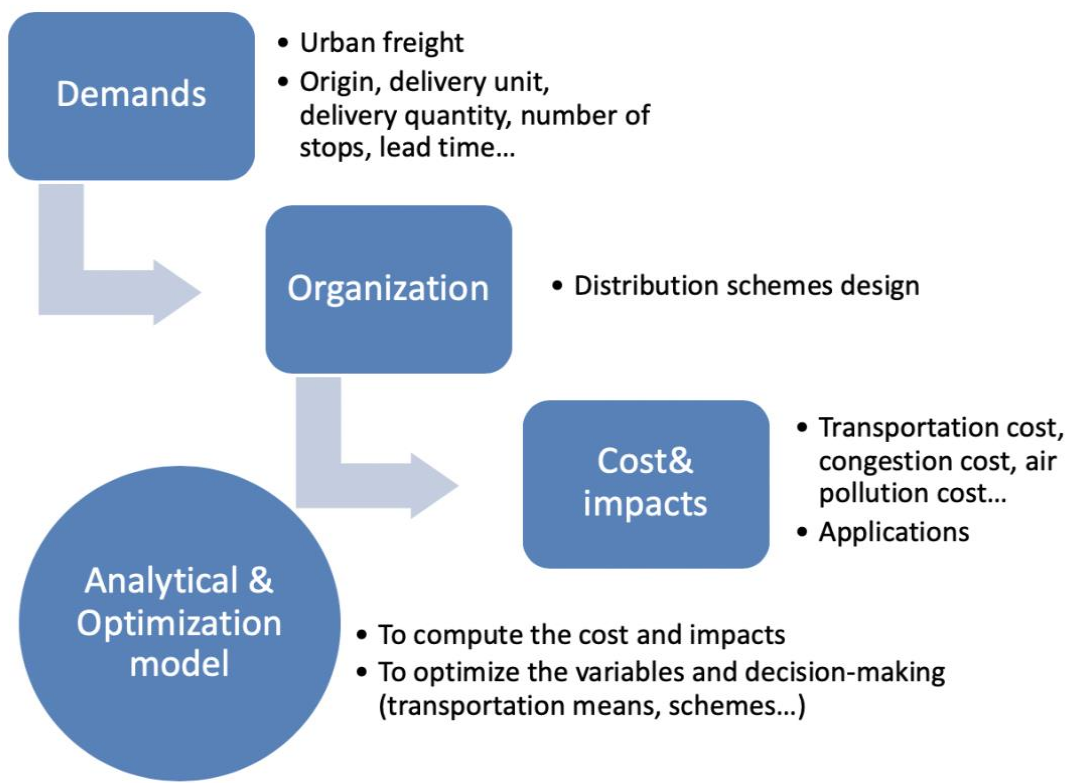

*Figure 1: Illustrating wholistic methodology of this study*

Figure 1 explains the methodology of our study, firstly input, following process, then output. The input should be demands including origin, delivery unit, delivery quantity, number of stops, lead time, transportation means used, and so on. The core of the process is the analytical and optimization models that translate the distribution schemes into mathematical equations and functions. These two models are used to compute the costs and external impacts, and optimize the variables (transportation means and distribution schemes), eventually to help decision-making. Lastly, the output is the results of costs and external impacts (accidents, air pollution, climate change, noise, and congestion), and further the applications of those results.

## 3.2   Analytical model

### 3.2.1   Problem description and assumptions

Based on the logistics cost models mentioned in the literature review, mainly the model of Kin et al. (2018), an analytical model to compute the transportation cost and external impacts is developed. There are certain types of vehicles used in the model. For each vehicle type $i$, the total number of tours $m$ is constrained by vehicle capacity, shift length and lead time at the same time. The total distance of the vehicle type $i$ is then calculated according to Daganzo (2005) method. Total transportation time includes travelling time and also the stop time. With the unit distance cost and unit time cost, the total distance cost and total time cost are obtained. The external impacts costs are calculated according to the parameters from the reports mentioned in the literature review.

Some assumptions are listed below:

- Lead time required by the customer is the same for each delivery;
- Stop time of each vehicle type at each stop is the same;
- One delivery for each stop in each tour;
- Each tour of the same vehicle type has the same delivery weight and number of stops;
- Each supplier delivers independently.

### 3.2.2   Notations

The notations of the analytical model in this subsection and also the optimization model in subsection 3.3 are listed below.

| Sets | |
|---|---|
| $i$ | Vehicle type |
| $j$ | Delivery unit type |
| **Parameters** | |
| $c_i$ | Capacity of vehicle type $i$ |
| $v_i$ | Average speed of vehicle type $i$ |
| $cd_i$ | Operation cost for vehicle type $i$ |
| $ct_i$ | Hourly wage for drivers of vehicle type $i$ |
| $w_j$ | Average weight of each stop in delivery unit type $j$ |
| $W$ | Total delivery weight per day |
| $Ns_j$ | Number of stops of delivery unit type $j$ |
| $Nss$ | Total number of stops |
| $\beta$ | Congestion factor |
| $tstop$ | Stop time at each stop |
| $sd$ | Shift duration |
| $lt$ | Lead time required by customers |
| $r$ | Average distance from supplier to PoS |
| $k$ | Factor in Daganzo method |
| $A$ | Area of the city |
| $p$ | Factor of accidents, air pollution, climate change, noise, and congestion according to v.km |
| *Variable* | |
| $np_{ji}$ | Percentage of delivery unit type $j$ allocated to vehicle type $i$ |

### 3.2.3 Cost function description

According to the notations, the number of the tours can be calculated as follows:

$$m_i = max\left(\left\lceil\frac{Ns_s \cdot W}{Ci}\right\rceil, \left\lceil\frac{\frac{di}{vi/\beta}+tstop \cdot Ns_s}{sd}\right\rceil, \left\lceil\frac{\frac{di-r}{vi/\beta}+tstop \cdot (Ns_s-1)}{lt}\right\rceil\right) \quad (1)$$

Since the number of tours $m_i$ should respect the vehicle capacity constraint, the shift length constraint (the maximum working hours of the driver), and the lead-time constraint (required by the customer), the maximum value of $m_i$ is considered.

$$d_i = 2 \cdot r \cdot m_i + k \cdot \sqrt{A \cdot Ns_s} \quad (2)$$

The travel distance of each tour is two times the distance from supplier to the first stop plus the distance between stops. Thus, the total distance $d_i$ in equation (2) equals to the number of tours multiply two times of the distance from supplier to the first stop plus the average distance between stops multiply the total number of stops minus one. As $m_i$ also depends on $d_i$, a recursive definition of $m_i$ is conducted.

$$Ctd = \sum_{i=1}^{n} di \cdot cdi \quad (3)$$

Total distance cost $Ctd$ is total distance multiply unit distance cost.

$$Ctt = \sum_{i=1}^{n} \left(\frac{di}{vi/\beta} + tstop \cdot Ns_s\right) \cdot cti \quad (4)$$

The total time $Ctt$ includes travel time and stop time. The total time cost is total time multiply unit time cost.

$$TC = Ctd + Ctt \quad (5)$$

Total transportation cost $TC$ is the summation of total distance cost and total time cost.

$$EIC = p \cdot \sum_{i=1}^{n} di \quad (6)$$

External impacts $EIC$ equal to constant $p$ multiply total v.km.

## 3.3 Optimization model: transportation means selection

In order to optimize transportation means selection, an optimization model is developed as followed.

The only decision variable $np_{ji}$ is the percentage of delivery unit type $j$ allocated to vehicle type $i$. For example, if vehicle type $i$ refers to van, light truck, heavy truck, delivery unit type $j$ refers to parcel, pallet and roll, then $np_{23}$ means the percentage of pallet allocated to heavy truck. The objective function $F$ is to minimize total transportation cost and the constraints satisfy vehicle capacity, shift duration and lead time.

$$\min \quad F = (\sum_i (cd_i \cdot (2 \cdot m_i \cdot r + k \cdot \sqrt{A \cdot \left(\frac{\sum_j np_{ji}}{Q}\right)}))$$
$$+ \sum_i \left( \left( \frac{2 \cdot m_i \cdot r + k \cdot \sqrt{A \cdot \left(\frac{\sum_j np_{ji}}{Q}\right)}}{\frac{v_i}{\beta}} \right) + tstop \cdot Ns \right) \cdot ct_i \right) \qquad (7)$$

*S.t.*

$$\sum_j w_j \cdot Ns_j \cdot np_{ji} \leq m_i \cdot c_i \qquad \forall i \qquad (8)$$

$$\frac{2 \cdot m_i \cdot r + k \cdot \sqrt{A \cdot \sum_j Ns_j \cdot np_{ji}}}{\frac{v_i}{\beta}} + tstop \cdot \sum_j Ns_j \cdot np_{ji} \leq sd \qquad \forall i \qquad (9)$$

$$(2 \cdot m_i \cdot r + k \cdot \sqrt{A \cdot \sum_j Ns_j \cdot np_{ji}} - r \cdot m_i)/\frac{v_i}{\beta} + tstop \cdot \sum_j Ns_j \cdot np_{ji} \qquad \forall i \qquad (10)$$
$$\leq m_i \cdot lt$$

$$\sum_i np_{ji} = 1 \qquad \forall j \qquad (11)$$
$$0 \leq np_{ji} \leq 1 \qquad \forall i,j \qquad (12)$$

Equation (7) is the objective function, which minimizes the total transportation cost (total distance cost plus total time cost). Since we focus on the organization level, the detailed vehicle routing problem (VRP) of each tour is not applied in our model. Instead, we assume that every tour of the same vehicle type is the same in terms of total delivery weight and the number of stops. In this way, the vehicle capacity, shift duration and lead time constraints are expressed in an overall view. Inequation (8) refers to vehicle capacity constraint, the total weight of delivery of vehicle type $i$ should not exceed the capacity of vehicle multiplying number of tours. Inequation (9) is the shift duration constraint, ensuring the total travel time and stop time of all tours is within the shift duration of the vehicle driver. Inequation (10) presents the lead time constraint, to finish delivering the last customer of the tour within lead time. Equation (11) ensures all the demand will be delivered. Inequation (12) is to define the decision variable as a percentage. Due to its none linearity and recursive definition of tours number, the model is optimized with a simulated annealing metaheuristic.

# 4   Case study

## 4.1 Data description

The delivery data of six suppliers from different sectors within 6 weeks in Bordeaux, France is obtained. There are totally 107 Point of Sales (PoS) with an average daily demand of 4215 kg of different goods in Bordeaux city and suburb area. And 8 vehicle types are used in terms of capacity (payload here): 2.3t, 8.1t, 9.45t, 10t, 12.15t, 14.85t, 17t, 25t. There are in total 5 types of goods in terms of temperature control namely A (normal temperature), F (fresh), S (frozen), T (three temperature), U (special temperature), so there are accordingly 5 types of vehicles in different temperature control with a distance unit cost of 5, 7, 8, 8, 5 €/km. Three delivery unit types are concluded, parcel (average weight of 10 kg), pallet (450kg), and roll (180kg). In each tour, there is only one delivery for each stop. And each supplier delivers independently.

There are three situations of temperature control:

- Normal vehicles are used for suppliers who only have dry goods at room temperature.
- Freezer vehicles are applied to the suppliers who only offer frozen goods.
- Containers with temperature control in normal vehicles are applied to the suppliers who have both normal goods and goods requiring different temperature control.

The delivery data are described in Table 2 and Table 3. We take an average day to analyze the *Key Performance Indicators (KPIs)* including total distance, total time spent (travel time and stop time), total transportation cost and fill rate of the vehicle when departure.

*Table 2: General description of the data input*

| | Total number of PoS | Total number of tours | Total number of deliveries | Average PoS per tour | Average weight per tour (kg) | Temperature control type | Vehicle type (in capacity) |
|---|---|---|---|---|---|---|---|
| Supplier 1 | 10 | 87 | 199 | 2.3 | 3043 | F | 17t, 25t |
| Supplier 2 | 42 | 767 | 1430 | 1.9 | 12943 | A, F, S | 25t |
| Supplier 3 | 12 | 82 | 240 | 2.9 | 2752 | A, F, S, U | 2.3t, 17t |
| Supplier 4 | 6 | 234 | 270 | 1.2 | 7345 | T | 8.1t, 9.45t, 12.15t, 14.85t |
| Supplier 5 | 29 | 134 | 715 | 5.0 | 5293 | S | 10t |
| Supplier 6 | 7 | 47 | 136 | 2.9 | 2764 | A | 10t |

*Table 3: Delivery data in an average day and results from the model*

| | Total deliveries | Total number of tours | Average number of stops per tour | Total weight (kg) | Total distance (km) | Total time (h) | Total cost (€) | Average fill rate in weight (%) |
|---|---|---|---|---|---|---|---|---|
| Supplier 1 | 6 | 2.4 | 2.3 | 7270 | 385 | 21 | 3040 | 26.4 |
| Supplier 2 | 36 | 19.2 | 1.9 | 248397 | 3102 | 164 | 29744 | 35.9 |
| Supplier 3 | 7 | 2.3 | 2.9 | 16435 | 280 | 16 | 1689 | 52.3 |
| Supplier 4 | 8 | 6.5 | 1.2 | 81000 | 286 | 16 | 1512 | 55.2 |
| Supplier 5 | 22 | 4.3 | 5.0 | 20590 | 1979 | 105 | 11052 | 15.1 |
| Supplier 6 | 5 | 1.7 | 2.9 | 4640 | 184 | 10 | 1047 | 23.9 |

## 4.2 Distribution schemes

To start with, three logistics schemes are studied: original, UCC, Simple PI.

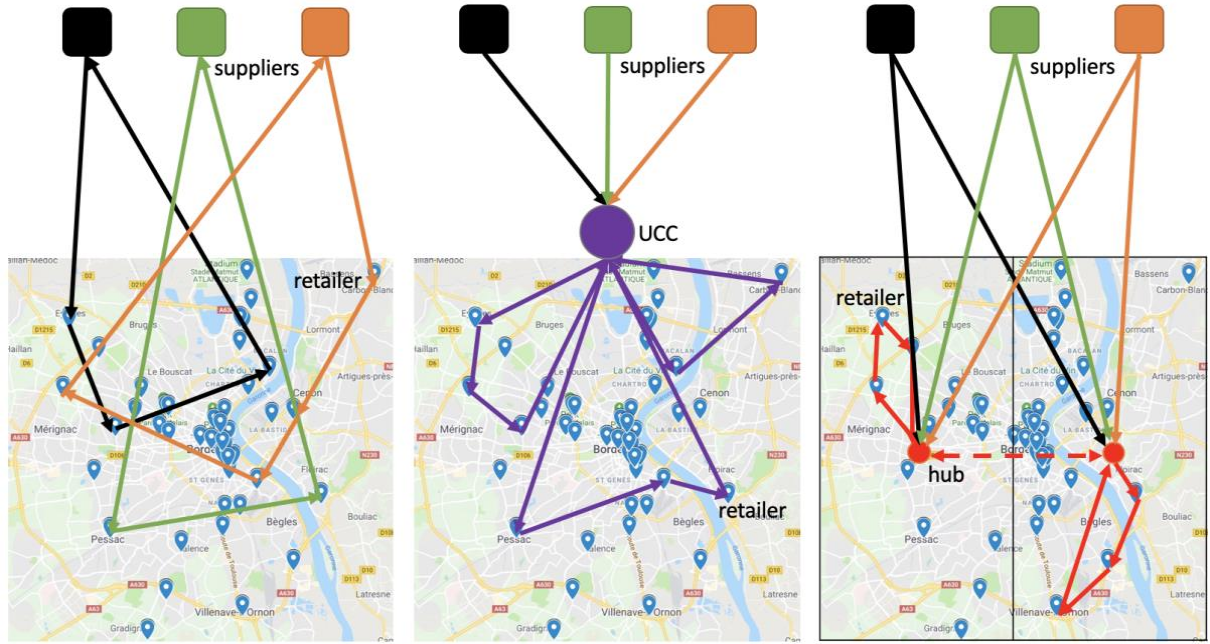

*Figure 2: Illustration of three schemes: original, UCC, simple PI.*

- Original scheme.

The original scheme serves as a benchmark to be compared with other schemes. In this scheme, six suppliers deliver independently. The detailed statistics are shown in Table 2 of subsection 4.1. All three situations of temperature control are applied. As is shown in Figure 2, the black, green and orange squares represent suppliers, and the blue drop shapes are the PoS. In the independent delivering context, the distribution routes of six suppliers might be partly overlapped, because each supplier can only deliver its own stops. The routes in black, orange, green represent routes from three different suppliers.

- UCC scheme.

The second one gathers all the demand of six suppliers to an Urban Consolidation Center (UCC) and arranges the delivery in a collaborative way. That is to say, firstly all the suppliers transport their goods to the UCC, then all demands are delivered from the UCC to PoS. For the first layer from suppliers to UCC, the situations of temperature control are as the same as the original scheme. When delivering goods from UCC to PoS, different vehicles with temperature control are applied: normal vehicles for goods at normal temperature, refrigerated vehicles for fresh or frozen goods.

We assume that the 40T truck (capacity of 25t) is used to deliver from suppliers to UCC due to the large quantity of each supplier. And 26T truck (capacity of 17t) is used from UCC to PoS, except 7.5T truck (2.3t) is used for goods with U temperature control because of its small quantity. So, in a vehicle tour of the second layer, there could be demands from PoS of different suppliers. For instance, the delivery tour in purple on the left includes stops from the black supplier and orange supplier in Figure 2. An additional handling cost at the UCC is also included in the total cost as 10€ per delivery, in which a delivery means all the parcels, pallets or rolls delivered to each stop in a single tour.

- Simple PI scheme.

The third one is to add two hubs (red points in Figure 2) which serve as a transshipment point in two subregions. The situation of temperature control and vehicle types are the same as the second scheme (UCC scheme). Firstly, each supplier delivers goods to each hub by heavy trucks (capacity of 25t), and we assume that there is a round trip for each hub. Then trucks (capacity of 17t) are used to deliver the PoS from the hub in each subregion. Referred to the definition of Physical Internet, transportation between hubs is possible when necessary. However, since the delivery quantity to each hub is scheduled according to the demand of PoS in each subregion, the transportation between hubs is not necessary in this case. The optimization of vehicle types used in the second layer from hubs to stops is investigated as well. Simple PI scheme with small vehicles (capacity of 2.3t) of the last layer is also investigated.

The analytical model is applied to all the three schemes with some different parameters (Table 4). To emphasize, for the UCC scheme and the simple PI scheme with hubs, the analytical model applies in the second layer, from UCC/hubs to PoS, so the distance $r$ has been changed to 10km/5km, and the area $A$ for the third scheme has been changed to 186/2 due to 2 subregions. When computing the KPIs, the results of the second layer should be the results of one sub region multiply 2. As for the first layer, from supplier to UCC/hubs, the distance is 20km/30 km and 25t truck is used. We assume the trucks make a round trip to UCC/each hub, based on the demand data, each hub needs two tours of the 25t truck. Since it happens in the suburb, the vehicle speed is higher as 30 km/h. Besides, an additional handling cost in the UCC/hubs is 10€ per delivery. That is basically how we compute the KPIs of the scheme with UCC/hubs.

*Table 4: Parameters in the analytical model used for three distribution schemes*

|  | Vehicle capacity $c_i$ (kg) | Vehicle speed $v_i$ (km/h) | Stop time $tstop$ (h) | Distance from supplier to stops $r$ (km) | Area $A$ (km2) |
|---|---|---|---|---|---|
| Original scheme | 2.3t, 8.1t, 9.45t, 10t, 12.15t, 14.85t, 17t, 25t | 20 | 0.25 | 30 | 186 |
| UCC scheme | 25t, 17t | 30, 20 | 0.5, 0.25 | 20, 10 | 186 |
| Simple PI scheme | 25t, 17t | 30, 20 | 0.5, 0.25 | 30, 5 | 186/2 |
| Simple PI scheme (small vehicle) | 25t, 2.3t | 30, 20 | 0.5, 0.25 | 30, 5 | 186/2 |

## 4.3 Sensitivity study

Investigating the impacts of some factors on the distribution performances is another concern of this study. These factors include vehicle speed and lead time. Obviously, with higher speed, it takes less time to travel the same distance, which will reduce the transportation cost. And with shorter lead time, frequent small trips delivering fewer stops are more practical than longer trips to deliver more stops, which will raise the transportation cost.

To validate the hypothesis about the impacts of vehicle speed and lead time, transportation cost and external impacts are evaluated under different values if vehicle speed and lead time. Results are presented in the next section.

# 5   Results

## 5.1   Distribution schemes

In this section, the results of different distribution schemes described in the previous section are analyzed. Some key performance indicators (KPIs) are calculated to evaluate different distribution schemes.

- Total distance refers to all the distances traveled by vehicles to deliver the total demand of one average day.
- Total time spent includes the travel time and stop time at each stop or each hub.
- Fill rate here is defined as the proportion of loaded goods in the vehicle capacity in terms of weight at departure. In order not to exceed the volume capacity, the weight of the maximum number of pallets/rolls is regarded as the actual capacity. For instance, the capacity of a 26T truck is 17t, but it can only take $30m^2$ of goods, which is 30 pallets ($1m^2$ per pallet). Therefore, the 26T truck cannot carry more than 7650 kg with 225kg per pallet. The actual capacity of a 26T truck is thus 7650 kg.
- Total cost is then the transportation cost including distance cost and time cost (plus handing cost for the scheme with hubs).

External impacts are illustrated separately in the next paragraph.

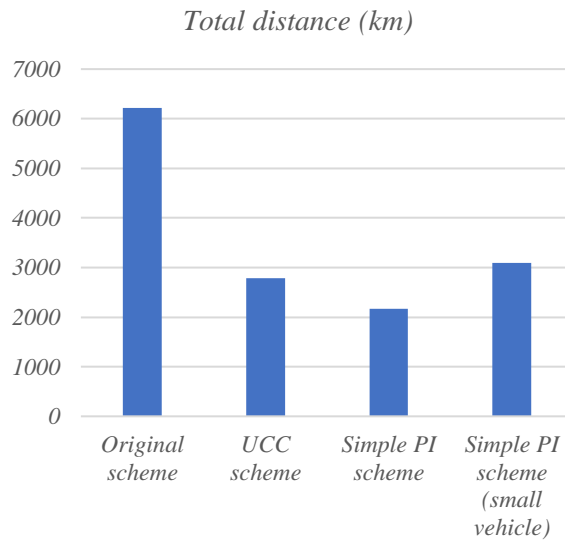
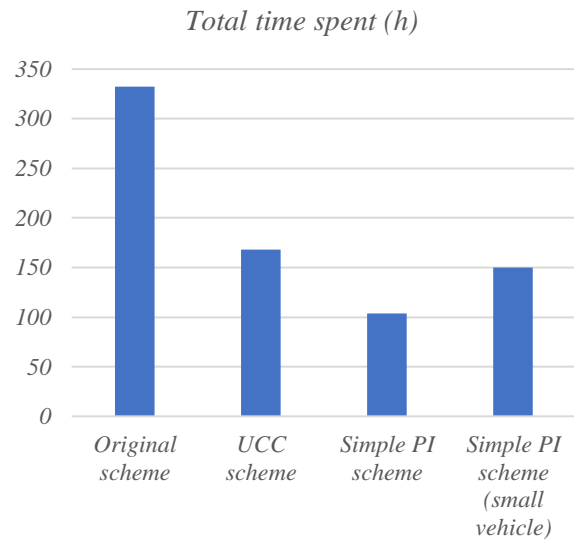

*Figure 3: Total Distance of Different Schemes*     *Figure 4: Total Time of Different Schemes*

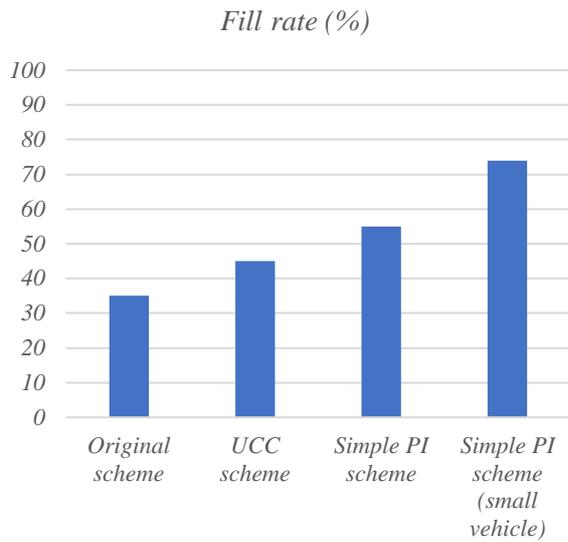

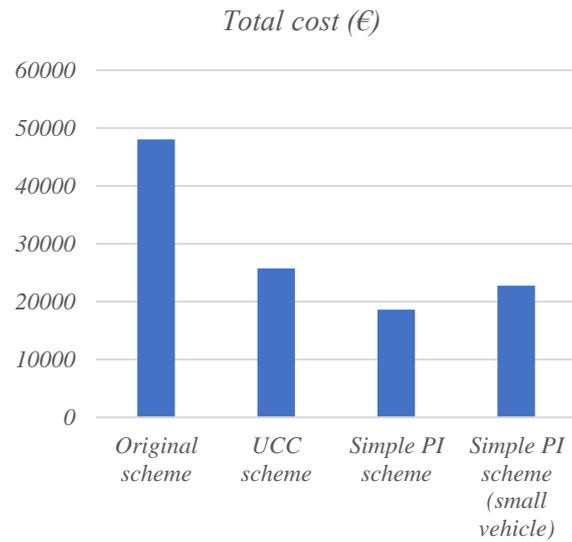

*Figure 5: Fill Rate of Different Schemes*     *Figure 6: Total Cost of Different Schemes*

As for external impacts, we apply the parameters of accidents, air pollution, climate change, and noise from INFRA/IWW report, the parameter of congestion from TU Delft report because there is no data about congestion in INFRA/IWW report. The results of the three schemes are shown in Figure 7 below.

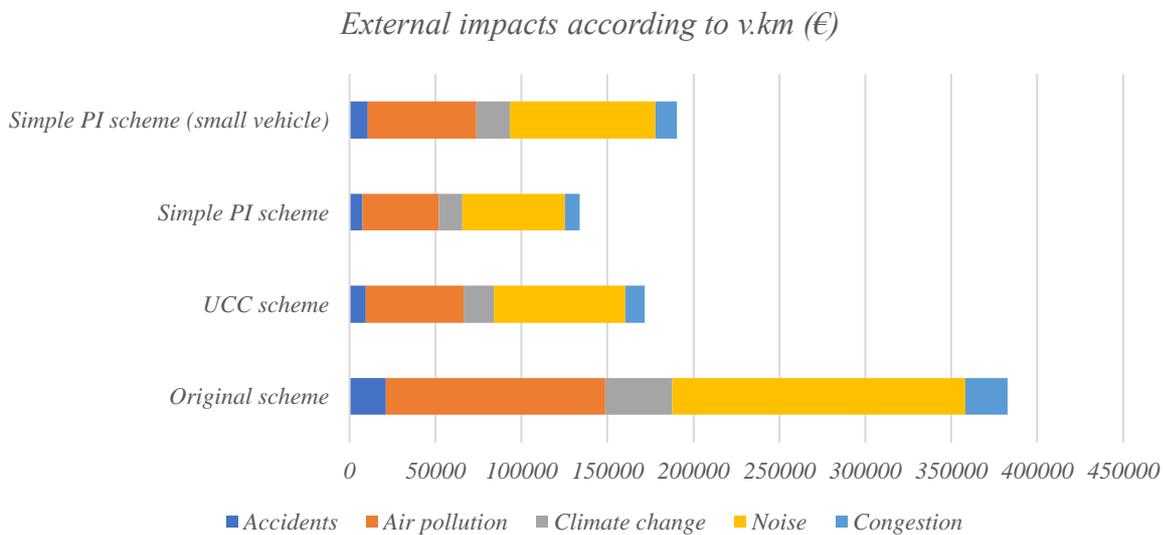

*Figure 7: External Impacts of Different Distribution Schemes*

## 5.2   Sensitivity study and optimization

To investigate the impacts of the variable (selection of transportation means) and parameters (vehicle speed, lead time), optimization and sensitivity study are conducted thanks to the analytical model.

The optimization model of transportation means selection in Section 3.3 is applied here. We have optimized the Simple PI scheme with small vehicles (Figure 8) and the original scheme of one single supplier (Figure 9) as well. It is meaningless to compare the original scheme of all the six suppliers with the optimization of them since they deliver independently but the optimization is conducted under the premise of all the suppliers collaboratively deliver together. Results are expressed in percentage to directly show the differences in vehicles choice optimization.

Figure 8 is the results of optimization compared with Simple PI scheme with small vehicles in Table 4. The application of vehicle types is the 25t truck from suppliers to hubs and 2.3t van from hubs to PoS before optimization. Vehicle types of the second layer from hubs to PoS are optimized among 25t, 17t and 2.3t trucks because they are respectively the largest, the most commonly used, and the smallest vehicles. The optimal solution, in this case, is to deliver all by the largest vehicle (25t). Although the fill rate has dropped 5% compared to the scheme without optimization, there is a significant decrease of the total distance (34%), total time (25%) and total cost (19%).

As for one single supplier in Figure 9, the original choice of vehicle types is the 25t truck for 10% of the goods and 17t truck for 90% of the goods. Among 25t, 17t and 2.3t vehicles, the optimization model minimizes total transportation cost with the final solution of delivering all by 17t truck. The improvements of KPIs are much more remarkable in this case, with a reduction of 42% of total distance, 38% of total time, 50% of total cost and an increase of 56% of fill rate.

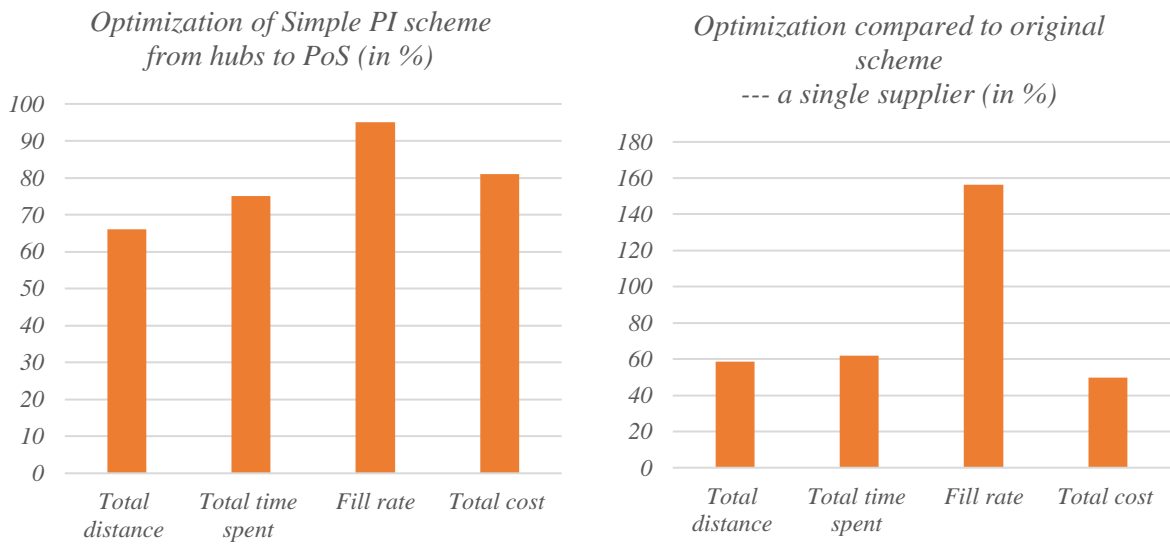

Figure 8: KPIs of Optimization of All Suppliers    Figure 9: KPIs of Optimization of One Supplier

The sensitivity study aims to investigate the impacts of vehicle speed and lead-time on distribution performances for the Simple PI scheme. Speed of the second layer from hubs to PoS is investigated, the increase of speed represents night delivery without congestion and the direct improvement of cost. The results are shown in Figures 10 and 11.

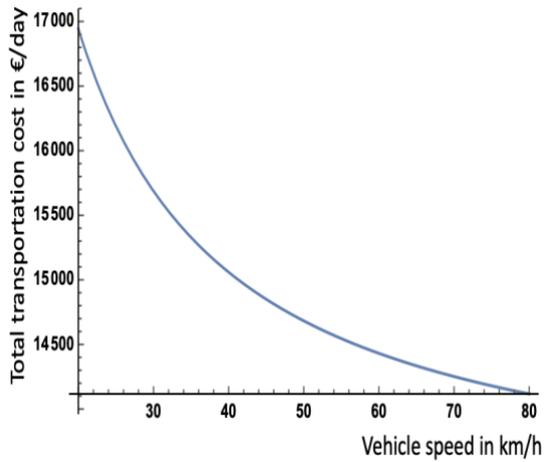

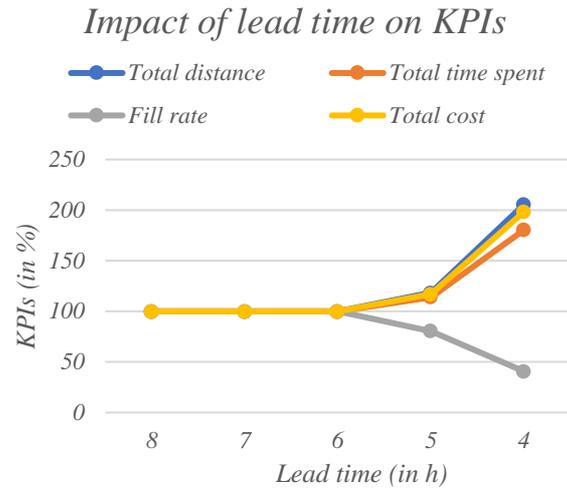

*Figure 10: Impact of Speed on Total Cost*  *Figure 11: Impact of Lead Time on KPIs*

In Figure 10, with the increase of vehicle speed, the total transportation decreases but total distance and fill rate are not changed. When we reduce the lead time from 8h to 4h in Figure 11, total distance, total time and total cost all remain the same at first and then rise at the point of 6h. The fill rate drops at the point of 6h. The results with a lead time lower than 4h cannot be calculated, because with the speed of 20 km/h, and the supplier 30 km far from PoS, it is impossible to meet all the demands within 4 hours.

## 6 Discussion

First of all, UCC and PI distribution both improve all the KPIs compared with the original scheme. Since there are some overlapped routes among the six suppliers, collaboration can effectively avoid them and reduce vehicle tours. With the same vehicle types, Simple PI scheme performs better than UCC scheme, it saves 28% of the total cost. Because the consolidation point is closer to the PoS, which results in a longer distance of the first layer but much smaller distance of the second layer. UCC reduces total distance more than simple PI scheme with small vehicles, due to more short vehicle tours. However, the utilization of smaller vehicles raises the fill rate from 45% (UCC) to 74% (Simple PI), which reduces total time as well as the total cost.

Secondly, external impacts have the same tendency as total distance, because they are calculated as a constant factor multiplying the total v.km. It implies that collaboration, both UCC and Simple PI, reduces significantly all these external impacts.

Finally, yet importantly, the impacts of transportation mean selection, vehicle speed and lead-time are studied. Thanks to optimization, a better selection of transportation means improve the KPIs in an overall view. The effectiveness of optimization differs from different scales of problem, for example, optimization reduces more cost and raises a higher fill rate in one single supplier than in all the six suppliers. And the optimal solution always avoids small vehicles but propose large vehicles (25t truck for the second layer of Simple PI scheme and 17t truck for one single supplier), because small vehicles results in much more tours (10 times by 2.3t truck as 25t truck for Simple PI scheme) for large scale of problems like this.

Vehicle speed doesn't affect the total distance and fill rate, but can reduce the total transportation cost by reducing the transportation time spent. It implies night delivery (with higher speed) could be a good solution. Shorter lead time at first doesn't change the distribution performances because it is relatively long. At a certain point (6h in this case study), when lead time becomes shorter, total distance, total time, and total cost all increase, while fill rate drops.

# 7  Conclusion

It is becoming increasingly challenging to deliver the city with lower cost and fewer external impacts. Great progress has been made in proposing solutions to this problem, and a literature review of possible solutions is conducted. In practice, there are lots of ongoing projects trying to evaluate those solutions. This paper investigates three different logistics schemes considering goods with temperature control to improve the efficiency and sustainability of urban logistics and provides insights for carriers as well as the policymakers of the city. Our contribution is based on an analytical and optimization models developed to evaluate the performances of different logistics schemes and perform sensitivity analysis. It turns out that optimization can improve the performances so the optimization model in this paper can be a tool for suppliers or carriers to optimize transportation means. Collaboration among suppliers can also improve distribution. Simple PI scheme performs better than UCC scheme. When applying small vehicles at the second layer, although it results in more short tours and longer distance, it reduces total time, total cost and improves fill rate. Simple PI scheme is thus regarded as the best logistics scheme among the three schemes studied here in this case study. Night delivery helps urban logistics operations, so it is recommended as an efficient solution. Furthermore, the limited time window is not beneficial for carriers or the environment.

There are still some limitations to this study. The first one is the optimization model only aims at minimizing the transportation cost. Another objective function to minimize external impacts should be added to have more sustainable solutions. Besides, transportation between hubs is not considered in this study. To further study the Physical Internet, a larger scale of case study including more hubs and transportation among hubs should be investigated.

Future research could improve the optimization model developed in this study by taking into consideration transportation means and hubs selection at the same time. The optimization of the number and location of hubs should be further considered. The optimization model could also be improved by adding another objective function of external impacts.

### Acknowledgement

The authors are grateful to Club Déméter Logistique Responsable itself and members for the collaboration during the case study and their useful comments.